\documentclass[letterpaper, 10 pt, conference, english]{ieeeconf} 

\IEEEoverridecommandlockouts  
\usepackage[utf8]{inputenc}

\usepackage{amsmath}
\usepackage{amsthm}
\usepackage{amssymb}
\usepackage{graphicx}

\makeatletter

\theoremstyle{plain}
\newtheorem{thm}{\protect\theoremname}
\theoremstyle{definition}
\newtheorem{problem}[thm]{\protect\problemname}
\theoremstyle{plain}
\newtheorem{prop}[thm]{\protect\propositionname}
\theoremstyle{plain}
\newtheorem{cor}[thm]{\protect\corollaryname}

\usepackage{bm}
\usepackage{hyperref}
\usepackage{xcolor}

\usepackage[normalem]{ulem} 

\makeatother

\usepackage{babel}

\providecommand{\corollaryname}{Corollary}
\providecommand{\problemname}{Problem}
\providecommand{\propositionname}{Proposition}
\providecommand{\theoremname}{Theorem}

\newif\ifshort

\shortfalse

\begin{document}

\title{\LARGE \bf Fleet Sizing in Vehicle Sharing Systems with Service Quality Guarantees}

\author{Michal \v{C}\'{a}p\thanks{Michal \v{C}\'{a}p is with FEL, CTU in Prague and TU Delft.}, 
Szabolcs Vajna\thanks{Szabolcs Vajna is with BUTE in Budapest.}, and 
Emilio Frazzoli\thanks{Emilio Frazzoli is with ETH Zurich and nuTonomy.}}

\maketitle
\begin{abstract}
Vehicle sharing system consists of a fleet of vehicles (usually bikes or cars) that can be rented at one station and returned at another station. We study how to achieve guaranteed
service availability in such systems. Specifically, we are interested
in determining a)~the fleet size and initial allocation of vehicles
to stations and b)~the minimum capacity of each station needed to
guarantee that a)~every customer will find an available vehicle at
the origin station and b)~the customer will find a free parking spot
at the destination station. We model the evolution of number of vehicles
at each station as a stochastic process and prove that the relevant
probabilities in the system can be approximated from above using a
computationally-tractable decoupled model. This property can be exploited
to efficiently determine the size of fleet, initial distribution of vehicles to stations, and station capacities
that are sufficient to achieve the desired service level. 
The applicability
of the method is demonstrated by computing the initial vehicle stock and the capacity of each station that would be
needed to avoid service failures in Boston's bike sharing system
``The Hubway". 
Our simulation shows that the proposed method is able to find more efficient design parameters than the naive approach and consequently it can achieve the equivalent quality-of-service level
with half of the vehicle fleet and half of the parking capacity.
\end{abstract}

\section{Introduction}

Bike sharing and car sharing systems can act as a
sustainable and economically viable alternative to private car ownership
in urban environments~\cite{shared-use_mobility_center_shared-use_2016}. 
A major obstacle to mass adoption of such vehicle
sharing systems is the low reliability of service \cite{katzev_richard_car_2003}. In
many such systems, the users regularly experience service unavailability:
Either there are no available vehicles at the origin station when they
decide to rent a vehicle or there is not enough parking spots at the destination
station when they decide to return a rented vehicle. In order to provide
customer experience that is comparable to the comfort of using a privately-owned
vehicle, a sharing system should \emph{guarantee} that a user will
always be able to pickup a vehicle at the desired origin location
and later return the vehicle at the destination location. Therefore,
a critical task of the system operator is to determine a) the size
of the vehicle fleet and the allocation of available vehicles to stations
at the beginning of the day, b) the capacity of each station, and c) a
strategy for future redistribution of vehicles between stations such
that no station becomes completely empty and
no station becomes completely full throughout the day. 

In recent years, researchers have increasingly focused on development of algorithms that support design and operation of vehicle sharing systems~\cite{gavalas_chapter_2016}.  

A large portion of existing work focuses on the analysis and synthesis of policies for
vehicle rebalancing -- since urban transportation patterns
are structurally imbalanced (e.g., in the morning, most customers
rent vehicles in residential areas and return them in business areas),
the vehicles must be continuously moved (rebalanced) from stations
with surplus of vehicles to stations with shortage of vehicles. 
In order to determine efficient rebalancing strategies and to quantify the amount of rebalancing needed in a particular vehicle sharing system, researchers proposed fluidic models~\cite{pavone_robotic_2012}, queue theoretic models~\cite{zhang_control_2016, calafiore_flow_2017} as well as different heuristics~\cite{spieser_shared-vehicle_2016}. 
The existing rebalancing models, however, neglect the issue of fleet sizing. 

Although the above methods can achieve balanced flows on expectation,
in practice, each station must have a sufficient stock of vehicles
and sufficient parking capacity to also cover the difference between
the incoming and outgoing flows when they deviate from the expectation. 

Spieser et al.~\cite{spieser_toward_2014} provide a method for determining a lower bound on a fleet size that is necessary to ensure passenger queue stability, i.e., waiting time should not grow to infinity. In practice, however, one is rather interested in ensuring high-level of availability, i.e., almost all passengers should be served immediately. George and Xia~\cite{george_fleet-sizing_2011-1} studied the relation between the fleet size and the service availability within a queue-theoretic model. Yet, the analysis is limited to systems with time-invariant demand in asymptotic regime when the number of vehicles in the system goes to infinity. Thus, in order to determine an appropriate fleet-size in a practical system, one typically resorts to simulation-based optimization~\cite{zhu_interplay_2017, zhu_toward_2017, fagnant_dynamic_2018}.  

The contribution of this paper is therefore a principled treatment
of the problem of determining the minimum initial fleet size and parking
capacity at each station needed to avoid service failures in the system
with desired confidence, given probabilistic time-dependent model of future demand
and a fixed rebalancing plan. To evaluate the probability of such
a service failure, we model the development of the number of vehicles
at each station as a stochastic process. Such a model would be prohibitively
expensive to evaluate, but we find that the relevant probabilities
can be approximated from above by a computationally-tractable model,
where the random variables that represent the number of vehicles at each station 
are assumed to be independent. 

This allows us to optimize over the initial number of vehicles at each station
and over the capacity at each station such
that system failures are avoided with required confidence. To demonstrate
the potential of the proposed method, we use the method to determine
the size of the fleet and capacities of stations necessary to avoid
service failures in Boston's bike sharing system with 99\% confidence.
Compared to a naive system design technique, the proposed method achieves the same level of service reliability with less than half of the vehicles and with less than half of the parking capacity. 

\section{\label{sec:Problem-Statement}Problem Statement}

Consider a vehicle sharing system consisting of $k$ stations, with
the set of all station labels denoted by $X=\{1,\ldots,k\}$. At time
$t=0$, station 1 stores $v_{1}$ vehicles, station 2 stores $v_{2}$
vehicles, etc. A vehicle moves from one station to another when either
a) a customer rents a vehicle at one station and returns the vehicle
at another stations or b) it is relocated as a part of the rebalancing
process. 

The customer demand is modeled probabilistically. In particular, the
requests to rent vehicle for transportation from station $o$ to station
$d$ are assumed to be generated by a nonhomogeneous Poisson point process
with intensity $\lambda_{od}(t)$ at time~$t$. The rebalancing vehicles
are relocated from station $o$ to station $d$ according to a deterministic
rebalancing plan. The plan for rebalancing between stations $o$ and
$d$ is represented as a set of time points $\rho_{od}=\{t_{od}^{1},t_{od}^{2},\ldots\}$
prescribing when rebalancing vehicles should depart from station $o$
towards station $d$. Further, let $\rho:=\cup_{o,d\in X}\ \rho_{od}$
denote the set of all time points when rebalancing occurs. Whenever
a rented or rebalancing vehicle departs from station $o$ destined
to station $d$, it will arrive to station $d$ with deterministic
delay $\eta_{ij}$. 

To successfully pickup a vehicle, the origin station must not be empty.
If all pickup requests are served, we say that the system is maintaining
full availability. And vice versa, if a customer requests a vehicle
and the station is empty, the system is said to fail to maintain full
availability. Similarly, each station $i$ has a limited parking capacity 
$c_{i}$, and a vehicle can be returned at the destination
station only when it is not at full capacity. When all arriving vehicles
find a free parking spot at their destination station, we say that
the system maintains capacity constraints. Conversely, the system
fails to maintain capacity constraints if a station is full and a
new vehicle arrives to the station. When the system maintains both
availability and capacity constraints we say that the system operates
failure-free.

As we can see, the failure-free operation of the system depends on
the initial stock of vehicles at each station, on the capacity of each station, and
on the rebalancing strategy. In this paper, we will assume that the vehicle sharing system employs a particular open-loop rebalancing strategy and we will focus on determining the initial number of vehicles and the
capacity of each station that collectively ensure failure-free operation of the
system. Since the future demand is generated by a Poisson process,
any station can experience arbitrarily high number of travel request
with non-zero probability. Therefore, failure-freeness of the system
cannot be achieved with certainty. It is, however, reasonable to ask
for a system that operates failure free with high probability as stated
in the following problem formulation:
\begin{problem}
\label{prob:determine_v_1_v_k}Determine the initial number of vehicles
$v_{1},\ldots,v_{k}$ that need to be present at respective stations
$1,\ldots,k$ at time $t=0$ and the capacity $c_{1},\ldots,c_{k}$
of each station $1,\ldots,k$ that ensure that the system will operate
failure-free in given future time interval $[0,T]$ with chosen confidence
level~$(1-z)$ with $z$ representing the maximum allowed probability of failure in the system. 
\end{problem}

\section{\label{sec:Coupled-Model}Coupled Model}

In this section, we will develop a numerical method for determining
the exact probability of failure-free operation of a vehicle sharing
system in the time interval of interest $[0,t]$. When developing
the model, we neglect the travel delay between stations to maintain
analytic tractability, but we show that it can be naturally incorporated in the simpler decoupled model of the next section. 

The evolution of the system is modeled as a continuous-time random
process $\{V(t)\}_{t\in[0,\infty)}$ over state space $\mathbb{M}\cup\{F\}$.
As long as the system operates failure-free, it remains in set $\mathbb{M}$,
once it experiences a failure, it falls into an absorbing state denoted
by $F$. The failure-free part of state space, defined as $\mathbb{M}:=\mathbb{M}_{1}\times\ldots\times\mathbb{M}_{k}$
with $\mathbb{M}_{i}:=\{0,\ldots,c_{i}\}$ being the set of all possible
numbers of vehicles at station $i$, contains a state for every possible
combination of number of vehicles at individual stations. Therefore,
if $V(t)\neq F$, then random variable $V(t)=(m_{1},\ldots,m_{k})$
represents the fact that the vehicle sharing system has $m_{1}$ vehicles
at station $1$, $m_{2}$ vehicles at station $2$, etc. To refer
to the number of vehicles at a single station, we define random
variables $V_{1}(t),\ldots,V_{k}(t)$ as 
\[
V_{i}(t):=\begin{cases}
(V(t))_{i} & \text{if }V(t)\in\mathbb{M}\\
F & \text{if }V(t)=F
\end{cases},
\]
where $(\mathbf{x})_{i}$ denotes the $i$-th element of tuple $\mathbf{x}$.
Observe that the random variables $V_{1}(t),\ldots,V_{k}(t)$ are
not independent. They are coupled, e.g., by the constraint
that the number of vehicles in the system is constant, i.e., for every
time point $t$ we have $\sum_{i=1,\ldots,k}V_{i}(t)=\sum_{i=1,\ldots,k}V_{i}(0)=\text{const}$.
Therefore, the model described in this section will be referred to
as coupled model. 

The probability that the system has been failure-free in time interval
$[0,t]$ and there is $m_{1}$ vehicles at station $1$, $m_{2}$
vehicles at station $2$, etc. at time $t$ is denoted by $p_{\mathbf{m}}(t)$,
where $\mathbf{m}=(m_{1},\ldots,m_{k})$ . We define $p_{\mathbf{m}}(t):=\mathbb{P}(V(t)=\mathbf{m})$
for any $\mathbf{m}\in\mathbb{M}$ and $p_{\mathbf{m}}(t):=0$ otherwise.

The collection of probabilities $p_{\mathbf{m}}(t')$ for all failure-free
states $\mathbf{m}\in\mathbb{M}$ and all time points $t'\in[0,t]$
can be expressed as a time-dependent $k$-dimensional tensor $\mathbf{P}(t')=\{p_{\mathbf{m}}(t')\}_{\mathbf{m}\in\mathbb{M}}\in[0,1]^{\mathbb{M}_{1}\times\ldots\times\mathbb{M}_{k}}$.
The probability that the system is failure-free in time interval $[0,t]$
is $p_{N}(t):=\,\mathbb{P}(V(t)\in\mathbb{M}).$
Similarly, the probability that the system has experienced a failure
in time interval $[0,t]$ is $p_{F}(t):=\mathbb{P}(V(t)=F).$
Observe that we have $p_{N}(t)=\sum_{\mathbf{m}\in\mathbb{M}}p_{\mathbf{m}}(t)$
and $p_{F}(t)=1-p_{N}(t).$ 

Further, we write $p_{F}(t;(v_{1},\ldots,v_{k}),(c_{1},\ldots,c_{k}))$
to denote the probability $\mathbb{P}(V(t)=F)$ when the initial numbers
of vehicles and capacities at stations $1,\ldots,k$ are fixed to
$v_{1},\ldots,v_{k}$ and $c_{1},\ldots,c_{k}$ respectively. 

In order to be able to concisely express the dynamics of the modeled
process, we need to introduce some additional notation. Firstly, the
following shorthand notation is used to refer to subsets of $\mathbb{M}$
consisting of states satisfying constraints on the number of vehicles
at some of the stations:
\[
\begin{array}{rl}
\mathbb{M}^{i=j}:= & \{\mathbf{m}|\mathbf{m}=(m_{1},\ldots,m_{k})\in\mathbb{M}\text{ and }m_{i}=j\},\\
\mathbb{M}^{i>j}:= & \{\mathbf{m}|\mathbf{m}=(m_{1},\ldots,m_{k})\in\mathbb{M}\text{ and }m_{i}>j\},\\
\mathbb{M}^{i<j}:= & \{\mathbf{m}|\mathbf{m}=(m_{1},\ldots,m_{k})\in\mathbb{M}\text{ and }m_{i}<j\}.
\end{array}
\]
Further, constraints may be chained, e.g., $\mathbb{M}^{i_{1}=j_{1},i_{2}<j_{2}}$
will be a shorthand for $\mathbb{M}^{i_{1}=j_{1}}\cap\mathbb{M}^{i_{2}<j_{2}}$. 

The family of functions $T_{od}:\ \mathbb{Z}^{k}\rightarrow\mathbb{Z}^{k}$
for $o,d\in X\times X$ encodes the effect of a relocation of single
vehicle from station $o$ to station $d$ on the number of vehicles
at each station. More specifically, $T_{od}(\mathbf{m})$ represents
the number of vehicles at each station when the system was in state
$\mathbf{m}$ and a single vehicle relocated from station $o$ to
station $d$: 
\begin{gather*}
T_{od} : \mathbf{m}=(m_{1},\ldots,m_{k})\mapsto\mathbf{m'}=(m'_{1},\ldots,m'_{k}),\\
\text{ where \ensuremath{m'_{i}=\begin{cases}
m_{i}-1 & \text{ if }i=o\\
m_{i}+1 & \text{ if }i=d\\
m_{i} & \text{otherwise.}
\end{cases}}}
\end{gather*}
As a special case, if the origin and destination stations are the same,
we define $T_{ii}(\mathbf{m})=\mathbf{m}$. Note that transformation $T$
does not account for capacity and availability constraints. Further,
let $T^{-1}$ denote the inverse of $T$, i.e., $(T_{od})^{-1}(\mathbf{m}=(m_{1},\ldots,m_{n}))=\mathbf{m'}\text{ such that }T_{od}(\mathbf{m}')=\mathbf{m}.$ 

We are now in position to describe the dynamics of the system. Suppose
that initially, at time $t=0$, the stations $1,\ldots k$ have $v_{1},\ldots,v_{k}$
vehicles respectively. Then, the initial condition of $p_{\mathbf{m}}$
is $p_{\mathbf{m}}(0)=1$ for $\mathbf{m}=(v_{1},\ldots,v_{k})$ and
$p_{\mathbf{m}}(0)=0$ for every other value of $\mathbf{m}$. The
system is initially failure-free and thus we have $p_{F}(0)=0.$ 

The functions $\{p_{\mathbf{m}}\}_{\mathbf{m}\in\mathbb{M}}$ and
$p_{F}$ are piecewise differentiable on $\mathbb{R}_{\geq0}$. At
time points $t_{1},t_{2},\ldots\in\rho$, vehicle is rebalanced between
stations and thus the functions $\{p_{\mathbf{m}}\}_{\mathbf{m}\in\mathbb{M}}$
and $p_{F}$ contain a jump discontinuity. Formally, for every $o,d\in X$,
and $t\in\rho_{od}$ we have 
\[
\begin{array}{rl}
p_{\mathbf{m}}(t)= & \lim\limits _{t'\rightarrow t^{-}}\ p_{T_{od}^{-1}(\mathbf{m})}(t')\quad\text{and }\\
p_{F}(t)= & \lim\limits_{t'\rightarrow t^{-}}\ p_{F}(t')+\sum_{\mathbf{m}\in\mathbb{M}^{o=0}\cup\mathbb{M}^{d=c_{i}}}p_{\mathbf{m}}(t').
\end{array}
\]

Everywhere else, i.e., at any $t\in\mathbb{R}_{\geq0}\setminus\rho$,
the functions $\{p_{\mathbf{m}}\}_{\mathbf{m}\in\mathbb{M}}$ and
$p_{F}$ are differentiable and their time evolution is governed by
the following system of differential equations. For every $\mathbf{m}\in\mathbb{M}$
and $t\in\mathbb{R}_{\geq0}\setminus\rho$, 
\[
\begin{aligned}\dot{p}_{\mathbf{m}}(t)= & \sum\limits _{o,d\in X}\lambda_{od}(t)\cdot p_{T_{od}^{-1}(\mathbf{m})}(t)-p_{\mathbf{m}}(t)\cdot\sum\limits _{o,d\in X}\lambda_{od}(t)\text{ and }\\
\dot{p}_{F}(t)= & \sum_{o,d\in X}\lambda_{od}(t)\sum_{\mathbf{m}\in\mathbb{M}^{o=0}\cup\mathbb{M}^{d=c_{i}}}p_{\mathbf{m}}(t).
\end{aligned}
\]
\begin{prop}
\label{prop:coupled-model-preserves-prob}This construction preserves
probability, that is, for every $t\in\mathbb{R}_{\geq0}$ we have
$\sum_{m\in\mathbb{M}}\ p_{\mathbf{m}}(t)+p_{F}(t)=1$. 
\end{prop}
\begin{proof}
\ifshort
See the appendix of on-line version~\cite{cap_fleet_arxiv_2018}. 
\else
See Appendix for formal proof. 
\fi
\end{proof}

\section{\label{sec:Decoupled-Method}Decoupled Method}

The evaluation of the coupled model requires numerical integration
over all elements from $\mathbb{M}$. Since the number of elements
in $\mathbb{M}$ grows exponentially in the number of stations, this
method becomes computationally intractable when applied to systems
with more than a few stations. 

For larger systems, we propose to use an alternative \emph{decoupled}
model, where the individual stations are modeled by independent stochastic
processes. In contrast to the coupled model, where the relocation
of vehicles between every pair of stations $o$ and $d$ is governed
by a single Poisson process with intensity $\lambda_{od}$, the relocation
of vehicles in the decoupled model is governed by a departure process
and an independent arrival process, both with intensity
$\lambda_{od}$. The departure process from $o$ to $d$ generates time points when
customers request to rent vehicles from station $o$ towards station
$d$, while the arrival process generates time points when vehicles
rented from station $o$ arrive to station $d$. Note that in such a setting,
the physical correspondence between vehicles from the departure process
and the vehicles from the arrival process is lost. In fact, it can happen that at some time point $t$, we
have more vehicles that have arrived at some station~$i$ than there are vehicles that have departed towards station~$i$. Analogously to the
coupled model, when a vehicle is requested from an empty station,
the station experiences an availability failure. When a vehicle arrives
to a full station, the station experiences capacity failure. 

Decoupling the departure and arrival processes enables us to reason
about the evolution of the number of vehicles at each station separately
and consequently evaluate the probability of failure more efficiently.
In the decoupled model, we model the evolution of the system as a
collection of $k$ random processes. The evolution of the stock of
vehicles at station $i$ is a continuous-time random process $\{\bar{V}_{i}(t)\}$
over state space $\mathbb{M}_{i}\cup\{F\}$. As long as the station
operates failure-free, it remains in the set $\mathbb{M}_{i}:=\{1,\ldots,c_{i}\}$,
once the station experiences a failure, it falls into an absorbing
state~$F$. Therefore, if $\bar{V}_{i}(t)\neq F$, then $\bar{V}_{i}(t)=j$
represents the fact that the station $i$ currently has $j$ vehicles. 

The probability that station $i$ at time $t$ has $j$ vehicles and
has not experienced a failure yet is denoted $q_{i}^{j}(t)$. We define
$q_{i}^{j}(t):=P(\bar{V}_{i}(t)=j)$ for $0\leq j\leq c_{i}$ and
$q_{i}^{j}(t):=0$ for other values of $j$. The probability that station $i$ has experienced a failure in time interval $[0,t]$ is
 $q_{i}^{F}(t):=P(\bar{V}_{i}(t)=F).$ 

Further, we write $q_{i}^{F}(t;v,c)$ to denote the probability $\mathbb{P}(\bar{V_{i}}(t)=F)$
when the initial number of vehicles at station~$i$ is $v$ and its
capacity is $c$. The initial condition for $q_{i}^{j}$ for every
$i\in X$ is $q_{i}^{j}(0)=1$ for $j=v_{i}$ , where $v_{i}$ is
the number of vehicles at station $i$ at time $0$, and $q_{i}^{j}(0)=0$
for other values of $j$. Initially, each station is failure-free
and thus we have $q_{i}^{F}(0)=0,\ \forall i\in X$. Let $\lambda_{i}^{a}(t):=\sum_{o\in X\setminus\{i\}}\lambda_{oi}(t)$
be the total intensity of all arriving vehicle flows to station~$i$
and $\lambda_{i}^{d}:=\sum_{d\in X\setminus\{i\}}\lambda_{id}(t)$
be the total intensity of all departing vehicle flows from station~$i$.
Analogously, let $\rho_{i}^{a}:=\cup_{o\in X\setminus\{i\}}\,\rho_{oi}$
be the set of time points when rebalancing vehicles are scheduled
to arrive to station~$i$ and $\rho_{i}^{d}:=\cup_{d\in X\setminus\{i\}}\,\rho_{id}$
be the set of time points when rebalancing vehicles are scheduled
to depart from station~$i$. The functions $\{q_{i}^{j}\}_{j\in\mathbb{M}_{i}}$
and $q_{i}^{F}$ for a particular station $i$ are piecewise differentiable
with a discontinuous jump at every time point $t\in\rho_{i}^{a}$
as follows
\[
q_{i}^{j}(t)=\lim\limits _{t'\rightarrow t^-}q_{i}^{j-1}(t');\quad q_{i}^{F}(t)=\lim\limits _{t'\rightarrow t^-}q_{i}^{F}(t')+q_{i}^{c_{i}}(t')
\]
and at every time point $t\in\rho_{i}^{d}$ as follows
\[
q_{i}^{j}(t)=\lim\limits _{t'\rightarrow t^-}q_{i}^{j+1}(t');\quad q_{i}^{F}(t)=\lim\limits _{t'\rightarrow t^-}q_{i}^{F}(t')+q_{i}^{0}(t').
\]
Everywhere else, i.e., for all $t\in\mathbb{R}_{\geq0}\setminus(\rho_{i}^{a}\cup\rho_{i}^{d})$,
the functions $\{q_{i}^{j}\}_{j\in\mathbb{M}_{i}}$ and $q_{i}^{F}$
are differentiable and their evolution is governed by the following
system of differential equations:
\begin{align*}
\dot{q}_{i}^{j}(t)= & \lambda_{i}^{a}\cdot q_{i}^{j-1}(t)-(\lambda_{i}^{a}+\lambda_{i}^{d})\cdot q_{i}^{j}(t)+\lambda_{i}^{d}\cdot q_{i}^{j+1}(t),\\
\dot{q}_{i}^{F}(t)= & \lambda_{i}^{a}(t)\cdot q_{i}^{c_{i}}+\lambda_{i}^{d}(t)\cdot q_{i}^{0}(t).
\end{align*}
If we denote $\mathbf{q}_{i}(t)=[q_{i}^{0}(t),\ldots,q_{i}^{c_{i}}(t)],$
then the behavior of the function $\mathbf{q}_{i}$ at discontinuous
jumps can be conveniently expressed as right or left ``shift'' of
the state vector: 
\[
\begin{aligned}\forall t\in\rho_{i}^{a}:\ \mathbf{q}_{i}(t)= & \lim\limits _{t'\rightarrow t^-}[0,q_{i}^{0}(t'),\ldots,q_{i}^{c_{i-1}}(t')]\\
= & [0,q_{0},\ldots,q_{c_{i}-1}]\text{, where }q=\lim\limits _{t'\rightarrow t^-}\mathbf{q}_{i}(t')\\
\forall t\in\rho_{i}^{d}:\ \mathbf{q}_{i}(t)= & \lim\limits _{t'\rightarrow t^-}[q_{i}^{1}(t'),\ldots,q_{i}^{c_{i}}(t'),0]\\
= & [q_{1},\ldots,q_{c_{i}},0]\text{, where }q=\lim\limits _{t'\rightarrow t^-}\mathbf{q}_{i}(t').
\end{aligned}
\]
The above system of differential equations can be expressed in
a matrix form $\dot{\mathbf{q}}_{i}(t)=Q_{i}(t)\cdot\mathbf{q}_{i}(t)$,
with $Q_{i}(t)=$ {\footnotesize{}
\begin{gather*}
\left[\begin{array}{ccccc}
-(\lambda_{i}^{d}+\lambda_{i}^{a}) & \lambda_{i}^{d} & 0 & 0 & \cdots\\
\lambda_{i}^{a} & -(\lambda_{i}^{d}+\lambda_{i}^{a}) & \lambda_{i}^{d} & 0 & \cdots\\
0 & \lambda_{i}^{a} & -(\lambda_{i}^{d}+\lambda_{i}^{a}) & \lambda_{i}^{d} & \cdots\\
0 & 0 & \lambda_{i}^{a} & -(\lambda_{i}^{d}+\lambda_{i}^{a}) & \cdots\\
\vdots & \vdots & \vdots & \vdots & \ddots
\end{array}\right].\\
\text{(the argument \ensuremath{t} of functions \ensuremath{\lambda_{i}^{d}} and \ensuremath{\lambda_{i}^{a}} was dropped for brevity)}
\end{gather*}
}In result, the vector function $\mathbf{q}_{i}$ can be efficiently
evaluated using a numerical computation software, and the probability
of failure at a station can be obtained as $q_{i}^{F}(t):=1-\sum_{j\in\mathbb{M}_{i}}q_{i}^{j}(t)$. 

We will now show that the decoupled model of the system can be used to obtain an upper-bounding approximation
of the failure probability in the coupled model.
\begin{thm}
\label{thm:decoupled-model-upperbounds-coupled}The probability that
a particular station $i$ has $j$ vehicles in the decoupled model
upper-bounds the probability that the same station $i$ has $j$ vehicles
in the coupled model, i.e., it holds that $P(V_{i}(t)=j)\leq P(\bar{V}{}_{i}(t)=j)$
for all $i\in X$, $j\in\mathbb{M}_{i}$, and $t\geq0$.
\end{thm}
\begin{proof}
\ifshort
See the appendix of on-line version~\cite{cap_fleet_arxiv_2018}. 
\else
See Appendix for formal proof. 
\fi
\end{proof}

For intuitive justification of this property, consider a system with three stations $X=\{a,b,c\}$ in which
at time $t$ we have an attempt to rent a vehicle from station~$a$ to station~$b$. 
In the decoupled model,
the arrival and departure processes are seen as independent and thus
the probability that station $b$ has, e.g., two vehicles after the
event, $P(\bar{V}_{b}(t)=2)$, is defined to be equal to the probability
that it had one vehicle before the event, ${P(\lim_{t'\rightarrow t^{-}}\bar{V}_{b}(t')=1)}$.
Notice that this probability does not depend on the state of station
$a$. In the coupled model, however, the relocation will not occur
if station $a$ is empty, which would be represented as $P(\lim_{t'\rightarrow t^{-}}V_{b}(t')=1\text{ and }V_{a}(t')\neq0)$
or equivalently $P(\lim_{t'\rightarrow t^{-}}V_{b}(t')=1)$ $-$ $P({\lim_{t'\rightarrow t^{-}}V_{a}(t')=0} \text{ and } V_{b}(t')=1)$.
Because the decoupled model does not account for the second term,
the probability $P(\bar{V}{}_{i}(t)=j)$ consistently overestimates
the probability $P(V_{i}(t)=j)$. Analogously, the probability that
station $c$ has, e.g., three vehicles after the event $P(\bar{V}_{c}(t)=3)$,
is in the decoupled model equal to $P(\lim_{t'\rightarrow t^{-}}\bar{V}_{c}(t')=3)$,
because there are no vehicles arriving to or departing from station~$c$.
In the coupled model, however, if station $a$ is empty or if station
$b$ is full, the entire system will fail and the probability of station
$c$ having three vehicles would be defined as $P(\lim_{t'\rightarrow t^{-}}V_{c}(t')=3\text{ and }V_{a}(t')\neq0\text{ and }V_{b}(t')\neq c_{b})$
or equivalently as $P({\lim_{t'\rightarrow t^{-}}V_{c}(t')=3})$ $-$
$P({\lim_{t'\rightarrow t^{-}}{V_{c}(t')=3} \text{ and } (V_{a}(t')=0}\text{ or }{V_{c}(t')=c_{d}}))$.
Again, the second term is not accounted for in the decoupled model
and thus the probability $P(\bar{V}_{c}(t)=3)$ overestimates probability $P(V_{c}(t)=3)$. 
Similar reasoning can be used to show that the sum
of station failure probabilities in the decoupled model also overestimates
the probability of system failure in the coupled model.

\begin{cor}
\label{cor:pf_is_bounded_by_sum_of_qif}The probability of failure
in the coupled model is upper bounded by the sum of probabilities
of failure at every station in the decoupled model, that is, $P(V(t)=F)\leq\sum_{i\in X}P(\bar{V}_{i}(t)=F)$. 
\end{cor}
\begin{proof}
\ifshort
See the appendix of on-line version~\cite{cap_fleet_arxiv_2018}. 
\else
See Appendix for formal proof. 
\fi
\end{proof}

Figure~\ref{fig:Coupled-decoupled-comparison} illustrates the discrepancy
between the failure probability in decoupled model $\sum_{i\in x}\:q_{i}^{F}$
and the failure probability in coupled model $p_{F}$ for an example
vehicle sharing system of practical size. We can see that quantity
$\sum_{i\in x}\:q_{i}^{F}$ is a reasonable approximation of $p_{F}$
for small values of $p_{F}$. Recall that our goal is to determine
if the probability of failure exceeds some small threshold value $z$,
e.g., $z=1\,\%$, at time point $T$, and thus we can make use of
this property and use $\sum_{i\in x}\:q_{i}^{F}$ as a substitute
of $p_{F}$.

\begin{figure}
\begin{centering}
\includegraphics[width=1\columnwidth]{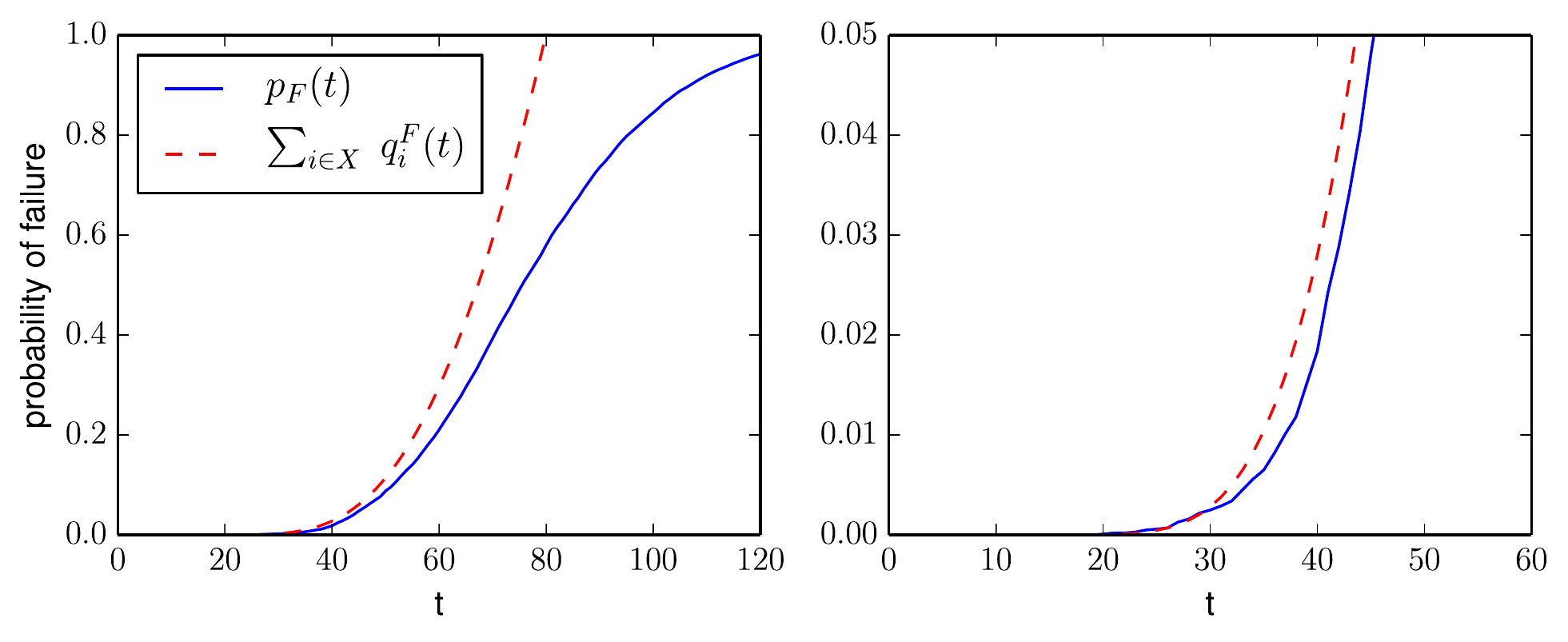}
\par\end{centering}
\caption{\label{fig:Coupled-decoupled-comparison}The illustration of the upper-bounding
approximation of the probability of failure obtained from the decoupled
model and the exact probability of failure in the coupled model. We
compare the two models using a vehicle sharing system with $k=50$ stations, where each
station $i$ has initially $v_{i}=50$ vehicles with maximum capacity
$c_{i}=100$ vehicles. The intensity of travel demand between every
two stations $o,d\in X$ is $\lambda_{od}=0.05$. The solid line shows
the probability of failure in the coupled model $p_{F}(t)$ obtained
by averaging 20000 simulated realizations of the random process, the
dashed line shows the upper bound on $p_{F}(t)$ obtained by evaluating
the decoupled model. The right plot shows the behavior of the two
functions for small values of $p_{F}(t)\leq5\%$. }

\end{figure}

The previous analysis assumed instantaneous relocation of vehicles
between stations, which allowed us to formulate the coupled model
as a memory-less random process and consequently maintain analytic tractability.
In the decoupled model, however, the travel delay can be incorporated
in a relatively straightforward way by appropriately shifting the
intensity of arrival process relative to the intensity of the respective
departure processes. That is, when the intensity of the departure
process from station $o$ to station $d$ at time $t$ is $\lambda_{od}(t)$,
then the intensity of the arrival process from station $o$ to station
$d$ at time $t$ is $\lambda_{od}^{a}(t):=\lambda_{od}(t-\eta_{od})$,
where $\eta_{od}$ is the travel time from station $o$ to station
$d$. Analogously, the total intensity of arrivals to station $i$
at time $t$ would be defined as $\lambda_{i}(t):=\sum_{o\in X}\ \lambda_{oi}(t-\eta_{oi})$
and the set of timepoints when rebalancing vehicles arrive to station
$i$ is $\rho_{i}^{a}:=\cup_{o\in X}\,\{t+\eta_{oi}|t\in\rho_{oi}\}$.

\section{Optimizing System Parameters~\label{sec:optimizing_vehicle_distribution}}

Now we return back to Problem~\ref{prob:determine_v_1_v_k}, which
asks for the number of vehicles needed initially at each station and
for the capacity of each station to ensure that the system will be failure
free in time interval $[0,T]$ with given given confidence level $1-z$.
Ideally, we would like to find the minimal number of vehicles and
the smallest number of parking spaces that suffice to satisfy our
quality of service constraint, i.e., we would like to solve the following
optimization problem 
\begin{equation}
\begin{aligned}\{(v_{i}^{\star},c_{i}^{\star})\}_{i}= & \underset{\{(v_{i},c_{i})\}_{i}\in\mathbb{N}^{2k}}{\mathrm{argmin}}\sum_{i}v_{i}+\sum_{i}c_{i}\\
 & \text{ subject to }p_{F}(T;\{(v_{i},c_{i})\}_{i})\leq z.
\end{aligned}
\label{eq:ideal_opt_prob}
\end{equation}
The evaluation of $p_{F}(T)$ is, however, computationally intractable for vehicle
systems of practical sizes. Instead, we can exploit Corollary~\ref{cor:pf_is_bounded_by_sum_of_qif}
and the observation that for small failure threshold $z$, the failure
probability in the coupled model can be bounded from above by
the sum of failure probabilities at each station to replace term $p_{F}(T;\{(v_{i},c_{i})\}_{i})$
with $\sum_{i}q_{i}^{F}(T;v_{i},c_{i})$, which yields the optimization
problem 
\begin{equation}
\begin{aligned}\{(v'_{i},c'_{i})\}_{i}= & \underset{\{(v_{i},c_{i})\}_{i}\in\mathbb{N}^{2k}}{\mathrm{argmin}}\sum_{i}v_{i}+\sum_{i}c_{i}\\
 & \text{ subject to }\sum_{i}q_{i}^{F}(T;v_{i},c_{i})\leq z.
\end{aligned}
\label{eq:coupled_opt_decoupled_constraint_problem}
\end{equation}
Note, that since $p_{F}(T;\{(v_{i},c_{i})\}_{i})\leq\sum_{i}q_{i}^{F}(T;v_{i},c_{i})$,
a solution$\{(v'_{i},c'_{i})\}$ to the optimization problem in Equation~\ref{eq:coupled_opt_decoupled_constraint_problem}
is also a feasible, albeit possibly suboptimal, solution for the problem
in Equation~\ref{eq:ideal_opt_prob}. The constraint function can
now be efficiently evaluated, but the formulation in Equation~\ref{eq:coupled_opt_decoupled_constraint_problem}
still represents a large-scale non-linear integer optimization problem,
which are notoriously hard to solve using existing methods. 

Observe that each individual term $q_{i}^{F}(T;v_{i},c_{i})$ in the
constraint function represents the probability that the system failure
occurs at station $i$. From this perspective, the optimization problem
must assign a budget of probabilities of failure to the individual
stations in a way that minimizes the total number of vehicles and
the total parking capacity used. This is again a challenging combinatorial
optimization problem. The complexity can be circumvented by picking
some desired partitioning of system failure probability $z$ to failure
probabilities at individual stations $z_{1},\ldots z_{k}$ such that
$z=\sum_{i}z_{i}$. For simplicity we will use uniform partitioning
that assigns $z_{i}=z/k$ for every station $i$. For any such fixed
partitioning, we can find the minimum number of vehicles $v_{i}$
and minimum parking capacity $c_{i}$ for each station $i$ as 
\[
v_{i},c_{i}=\underset{v,c\in\mathbb{N}}{\mathrm{argmin}}\ v+c\ \text{ subject to }q_{i}^{F}(T;v,c)\leq z_{i}.
\]

The above can be solved, e.g., by exhaustive enumeration of all value
combination for parameters $v$ and $c$. To improve performance,
we find an approximate optimal values for each station by optimizing
by coordinates. Recall that the evolution of $q_{i}^{F}$ is governed
by equation $\dot{q}_{i}^{F}(t)=\lambda_{i}^{a}(t)\cdot q_{i}^{c_{i}}+\lambda_{i}^{d}(t)\cdot q_{i}^{0}(t)$
with initial condition $q_{i}^{F}(0)=0$. The two terms in the differential
equation in fact represent the probability of capacity failure and
availability failure respectively. The probability of availability failure decreases with
increasing number of vehicles and the probability of capacity failure
decreases with increasing parking capacity at the station. Therefore,
we start by finding the minimal number of vehicles that suffices to
cap the probability of availability failure by $z_{i}/2$ as
\[
v_{i}=\underset{v\in\mathbb{N}}{\mathrm{argmin}}\ v\ \text{ subject to }q_{i}^{F}(T;v,\infty)\leq\frac{z_{i}}{2}.
\]
Then, for fixed $v_{i}$, we find the minimal parking capacity at
station $i$ that ensures that the probability of system failure at the station
is no more than $z_{i}$ as 
\[
c_{i}=\underset{c\in\mathbb{N}}{\mathrm{argmin}}\ c\ \text{ subject to }q_{i}^{F}(T;v_{i},c)\leq z_{i}.
\]
Since both $q_{i}^{F}(T;v,\infty)$ and $q_{i}^{F}(T;v_{i},c)$ are
monotonically non-increasing in $v$ and $c$ respectively, these
optimization problems can be efficiently solved, e.g., by the method
of bisection. The series of steps described above can find the
system parameters $\{(v_{i},c_{i})\}$ that represent a feasible solution
to problem in Equation~\ref{eq:ideal_opt_prob}. 

\section{Case Study: Hubway in Boston \label{sec:Boston_Case-Study}}

\begin{figure}
\begin{centering}
\hspace{-1mm}\includegraphics[width=5.5cm]{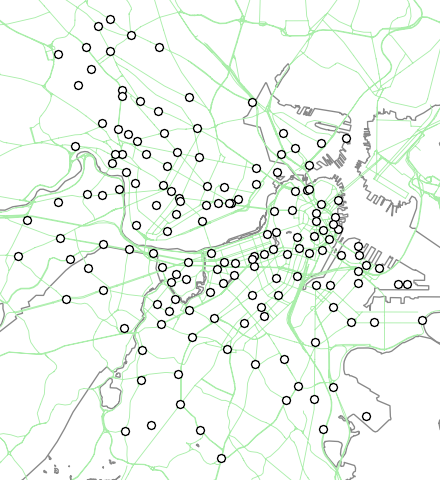}
\par\end{centering}
\caption{\label{fig:hubway-stations} The stations of Boston's bike sharing system ``The Hubway".}
\end{figure}

In this section, we demonstrate the applicability of the proposed
technique in the context of Boston's bike sharing system called ``The
Hubway''. This system consists of roughly 150 stations scattered
over the larger Boson area as shown in Figure~\ref{fig:hubway-stations}.  
The Hubway users often face availability
or capacity failures in the system. As a hypothetical exercise, we
use our method to determine what fleet size and parking capacity would
be sufficient to avoid such failure events. 

The Hubway releases anonymized historical data about all
bike rentals in the system. We use this dataset to estimate the parameters
of the system model and to evaluate performance of a particular system
design using historical rental data. For our experiment, we extracted
a data set of all bike rental records realized during working
days of May 2016. Then, we used the data set to estimate
the intensities of demand generating processes $\{\lambda_{od}\}$
in our model and to estimate the travel times $\{\eta_{od}\}$. 

\begin{figure}
\begin{centering}
\hspace{-1mm}\includegraphics[width=9.1cm]{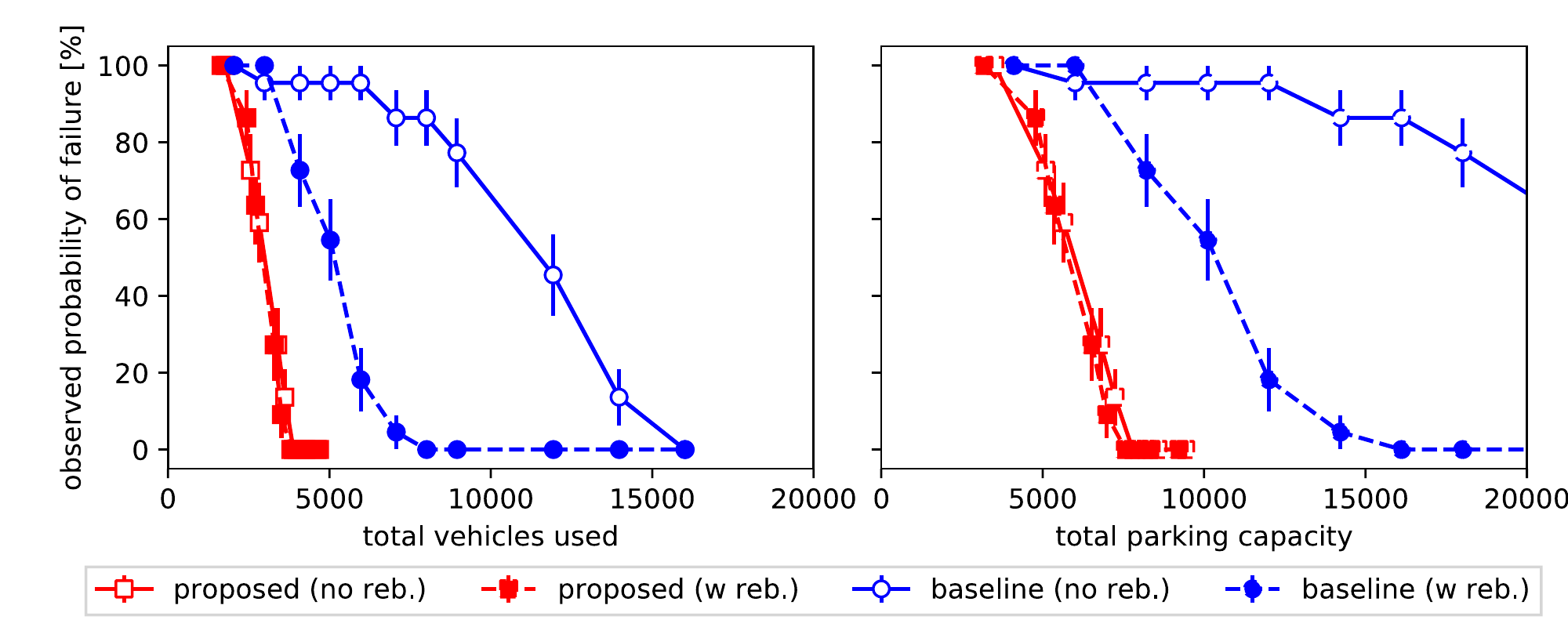}
\par\end{centering}
\caption{\label{fig:baseline-comparision-hubway}The plot shows the relation
between probability of failure during a day (24h period) and the total fleet size (left) and
total parking capacity used (right) when 1) each station has the same
parking capacity and initially it is half filled with bikes and half
empty (baseline) and 2) when the initial number of bikes and parking
capacity at each station is computed using the proposed method (proposed).
The dashed line shows the relation when rebalancing of bikes is used
and the solid line shows the relation for systems without rebalancing.}
\end{figure}

As expected, the structure of the demand is imbalanced. One can either
ignore the imbalance or try to counteract it by relocating vehicles
from stations where surplus of bikes is expected to stations where
shortage of bikes is expected. In our experiment, we consider both
systems that do not use rebalancing and systems where the vehicles
are rebalanced during the day. For systems with rebalancing, we find
an a priori rebalancing strategy that on expectation balances
the rate of bike arrivals and the rate of bike departures at each
station using the method proposed in \cite{spieser_toward_2014}. 

Next, we need to determine the initial number of vehicles at each
station and the parking capacity at each station that suffice to avoid
service failures with desired confidence. We compare the optimization-based method proposed
in Section~\ref{sec:optimizing_vehicle_distribution} with a baseline
method that assigns equal parking capacity to all stations and initially
fills half of the parking spaces at each station with bikes. Then,
we evaluate how a system with given design parameters would perform if
it had to face a particular sequence of rentals during a day. For performance evaluation, we use
22 historical rental sequences that correspond to rental demand during each working day of May 2016. 
For every rental
recorded in the sequence and for every rebalancing request dictated
by the rebalancing plan (if rebalancing is used), we relocate a single
vehicle from origin of the rental request to the destination with
appropriate travel delay. If the origin station is empty or the destination
station is full during any of such relocation, we record that the
system failed on that day. 

The plots in Figure~\ref{fig:baseline-comparision-hubway} show the
probability of system failure during one day as a function of fleet size and total
parking capacity for the proposed method and the baseline method,
both with and without rebalancing. We can observe that the proposed
method assigns the vehicles and parking capacity to stations more efficiently 
than the baseline approach. 
In result, by applying the proposed approach, one can avoid service failures in the system with overwhelming probability ($\geq 99\,\%$) using only half of the vehicle fleet and half of the parking capacity compared to what would be needed in a system designed using a baseline approach.

\section{Conclusion\label{sec:Conclusion}}
	
We studied if it is possible to design a vehicle sharing system with
service availability guarantees, i.e., a system where all passengers
are served with given, arbitrarily high probability. In particular,
we propose a systematic method for determining the appropriate system
design parameters, namely the initial number of vehicles at each station
and the parking capacity at each station, that are sufficient to guarantee
that a) all attempts to pickup a vehicle are successful and b) all
attempts to return a vehicle are successful, with a given minimum confidence
level. In order to determine the probability of a service failure
for a particular system design, one has to reason about stochastic
vehicle relocations and their effect on the stock of vehicles at each
station. However, the model of such system is a random process with exponentially
large state space that is prohibitively expensive to evaluate. 

Our main contribution is a theoretical analysis leading to an insight
that the relevant phenomena in such a system can be sufficiently grasped
in a simpler, decoupled stochastic model, where each station is analyzed
separately. We prove that the probability of failure in the decoupled
model is an upper bounding approximation of the failure probability
in the coupled model. Such a finding can be used to efficiently determine the
system design parameters that guarantee the desired quality of service
in the system. We demonstrate the applicability of the method by computing
the system design parameters for a bike sharing system in Boston,
and show that the proposed method can achieve the same quality of service
as baseline method with only half of the bike fleet and half of the parking capacity.

\paragraph*{Acknowledgments}

We would like to thank Slava Kungurtsev for assistance with some of
the mathematical problems that the theoretical analysis in this paper
hinges on. This   research   was   supported   by   the   Czech   Science
Foundation    (grant    No.    18-23623S) and OP VVV MEYS funded project CZ.02.1.01/0.0/0.0/16\_019/0000765 ``Research Center for Informatics". Access  to  computing  and  storage
facilities  owned  by  parties  and  projects  contributing  to  the
National  Grid  Infrastructure  MetaCentrum,  provided  under
the  program  ``Projects  of  Large  Infrastructure  for  Research,
Development,  and  Innovations"  (LM2010005),  is  greatly  appreciated.

\bibliographystyle{plain}
\bibliography{zotero}

\begin{thebibliography}{10}

\bibitem{calafiore_flow_2017}
G.~C. Calafiore, C.~Novara, F.~Portigliotti, and A.~Rizzo.
\newblock A flow optimization approach for the rebalancing of mobility on
  demand systems.
\newblock In {\em 2017 {{IEEE}} 56th {{Annual Conference}} on {{Decision}} and
  {{Control}} ({{CDC}})}, pages 5684--5689, December 2017.

\bibitem{fagnant_dynamic_2018}
Daniel~J. Fagnant and Kara~M. Kockelman.
\newblock Dynamic ride-sharing and fleet sizing for a system of shared
  autonomous vehicles in {{Austin}}, {{Texas}}.
\newblock {\em Transportation}, 45(1):143--158, January 2018.

\bibitem{gavalas_chapter_2016}
D.~Gavalas, C.~Konstantopoulos, and G.~Pantziou.
\newblock Chapter 13 - {{Design}} and management of vehicle-sharing systems: A
  survey of algorithmic approaches.
\newblock In Mohammad~S. Obaidat and Petros Nicopolitidis, editors, {\em Smart
  {{Cities}} and {{Homes}}}, pages 261--289. {Morgan Kaufmann}, Boston, 2016.

\bibitem{george_fleet-sizing_2011-1}
David~K. George and Cathy~H. Xia.
\newblock Fleet-sizing and service availability for a vehicle rental system via
  closed queueing networks.
\newblock {\em European Journal of Operational Research}, 211(1):198--207, May
  2011.

\bibitem{katzev_richard_car_2003}
{Katzev Richard}.
\newblock Car {{Sharing}}: {{A New Approach}} to {{Urban Transportation
  Problems}}.
\newblock {\em Analyses of Social Issues and Public Policy}, 3(1):65--86,
  December 2003.

\bibitem{pavone_robotic_2012}
Marco Pavone, Stephen~L Smith, Emilio Frazzoli, and Daniela Rus.
\newblock Robotic load balancing for mobility-on-demand systems.
\newblock {\em The International Journal of Robotics Research}, 31(7):839--854,
  June 2012.

\bibitem{spieser_shared-vehicle_2016}
Kevin Spieser, Samitha Samaranayake, Wolfgang Gruel, and Emilio Frazzoli.
\newblock Shared-{{Vehicle Mobility}}-on-{{Demand Systems}}: {{A Fleet
  Operator}}'s {{Guide}} to {{Rebalancing Empty Vehicles}}.
\newblock In {\em Transportation {{Research Board}} 95th {{Annual Meeting}}}.
  {Transportation Research Board}, 2016.

\bibitem{spieser_toward_2014}
Kevin Spieser, Kyle Treleaven, Rick Zhang, Emilio Frazzoli, Daniel Morton, and
  Marco Pavone.
\newblock Toward a {{Systematic Approach}} to the {{Design}} and {{Evaluation}}
  of {{Automated Mobility}}-on-{{Demand Systems}}: {{A Case Study}} in
  {{Singapore}}.
\newblock {\em Road Vehicle Automation (Lecture Notes in Mobility)}, April
  2014.

\bibitem{shared-use_mobility_center_shared-use_2016}
Shared use Mobility~Center.
\newblock Shared-use {{Mobility}} - {{Reference Guide}}.
\newblock Technical report, {Shared-use Mobility Center}, October 2016.

\bibitem{zhang_control_2016}
Rick Zhang and Marco Pavone.
\newblock Control of {{Robotic Mobility}}-on-demand {{Systems}}.
\newblock {\em Int. J. Rob. Res.}, 35(1-3):186--203, January 2016.

\bibitem{zhu_interplay_2017}
Shirley Zhu and Alain Kornhauser.
\newblock The {{Interplay Between Fleet Size}}, {{Level}}-of-{{Service}} and
  {{Empty Vehicle Repositioning Strategies}} in {{Large}}-{{Scale}},
  {{Shared}}-{{Ride Autonomous Taxi Mobility}}-on-{{Demand Scenarios}}.
\newblock In {\em Transportation {{Research Board}} 96th {{Annual Meeting}}},
  2017.

\bibitem{zhu_toward_2017}
Tong Zhu.
\newblock Toward a systematic approach to the fleet size estimation of
  autonomous mobility-on-demand systems.
\newblock Msc {{Thesis}}, {Cornell University}, May 2017.

\end{thebibliography}

\ifshort
\else
\section*{Appendix\label{sec:Appendix}}

\subsection*{Proof of Proposition~\ref{prop:coupled-model-preserves-prob}:}
\begin{proof}
Observe that for a particular $o,d\in X$, it holds that 
\[
\sum_{m\in\mathbb{M}}p_{T_{od}^{-1}(\mathbf{m})}(t)=\sum_{m\in\mathbb{M}^{o>1,d<c_{d}}}p_{\mathbf{m}}(t)
\]
and consequently we also have
\begin{multline*}
\sum_{m\in\mathbb{M}}\sum\limits _{o,d\in X}\lambda_{od}(t)\cdot p_{T_{od}^{-1}(\mathbf{m})}(t)\\
=\sum_{m\in\mathbb{M}^{o>1,d<c_{d}}}\sum\limits _{o,d\in X}\lambda_{od}(t)p_{\mathbf{m}}(t).
\end{multline*}

The sum $\sum_{m\in\mathbb{M}}\ p_{\mathbf{m}}(t)+p_{F}(t)$ is preserved
over every discontinuity $t\in\rho$. To see that 
\[
\sum_{m\in\mathbb{M}}\ p_{\mathbf{m}}(t)+p_{F}(t)=\lim\limits _{t'\rightarrow t^{-}}\sum_{m\in\mathbb{M}}\ p_{\mathbf{m}}(t')+p_{F}(t')
\]
 holds for any $t\in\rho$, take arbitrary $o,d\in X$ and $t\in\rho_{od}$
and perform the following rearrangement  
\begin{gather*}
\sum_{m\in\mathbb{M}}\ p_{\mathbf{m}}(t)+p_{F}(t)\\
=\sum_{m\in\mathbb{M}}\lim\limits _{t'\rightarrow t^{-}}\ p_{T_{od}^{-1}(\mathbf{m})}(t')\\
+\lim\limits _{t'\rightarrow t^{-}}\ p_{F}(t')+\sum_{\mathbf{m}\in\mathbb{M}^{o=0}\cup\mathbb{M}^{d=c_{i}}}p_{\mathbf{m}}(t')\\
=\lim\limits _{t'\rightarrow t^{-}}\sum_{m\in\mathbb{M}^{o>1,d<c_{d}}}p_{\mathbf{m}}(t')+p_{F}(t')+\sum_{\mathbf{m}\in\mathbb{M}^{o=0}\cup\mathbb{M}^{d=c_{i}}}p_{\mathbf{m}}(t')\\
=\lim\limits _{t'\rightarrow t^{-}}\sum_{m\in\mathbb{M}}\ p_{\mathbf{m}}(t')+p_{F}(t').
\end{gather*}

Further, we have $\sum_{m\in\mathbb{M}}\dot{p}_{\mathbf{m}}(t)-\dot{p}_{F}(t)=0$
at all $t\in\mathbb{R}_{\geq0}\setminus\rho$ by the following rearrangement
\[
\begin{aligned}\sum_{m\in\mathbb{M}}\dot{p}_{\mathbf{m}}(t)= & \sum\limits _{o,d\in X}\lambda_{od}(t)\left(\sum_{m\in\mathbb{M}^{o>1,d<c_{d}}}p_{\mathbf{m}}(t)-\sum_{m\in\mathbb{M}}p_{\mathbf{m}}(t)\right)\\
= & -\sum\limits _{o,d\in X}\lambda_{od}(t)\sum_{\mathbf{m}\in\mathbb{M}^{o=0}\cup\mathbb{M}^{d=c_{i}}}p_{\mathbf{m}}(t)\\
= & -\dot{p}_{F}(t),
\end{aligned}
\]
and thus the sum $\sum_{m\in\mathbb{M}}\ p_{\mathbf{m}}(t)+p_{F}(t)$
is also preserved over every interval where functions $\{p_{\mathbf{m}}\}_{\mathbf{m}\in\mathbb{M}}$
and $p_{F}$ are differentiable. 

We have $\sum_{m\in\mathbb{M}}\ p_{\mathbf{m}}(0)+p_{F}(0)=1$ at
$t=0$ from the initial conditions and since $\sum_{m\in\mathbb{M}}\ p_{\mathbf{m}}(t)+p_{F}(t)$
is constant for all $t>0$, we conclude that $\sum_{m\in\mathbb{M}}\ p_{\mathbf{m}}(t)+p_{F}(t)=1$
for all $t\geq0$. 
\end{proof}

\subsection*{Proof of Theorem~\ref{thm:decoupled-model-upperbounds-coupled}:}
\begin{proof}
Let $s_{i}^{j}(t):=P(V_{i}(t))=\sum_{\mathbf{m}\in\mathbb{M}^{i=j}}p_{\mathbf{m}}(t)$.
Define
\begin{gather*}
X_{od}^{i=j}(t):=\sum_{\mathbf{m}\in\mathbb{M}^{i=j}}p_{T_{od}^{-1}(\mathbf{m})}(t)\\
=\begin{cases}
\sum_{\mathbf{m}\in\mathbb{M}^{i=j,o>0,d<c_{d}}}p_{\mathbf{m}}(t) & \text{ if }o\neq i\text{ and }d\neq i\\
\sum_{\mathbf{m}\in\mathbb{M}^{i=j-1,o>0}}p_{\mathbf{m}}(t) & \text{ if }d=i\text{ and }j>1\\
\sum_{\mathbf{m}\in\mathbb{M}^{i=j+1,d<c_{d}}}p_{\mathbf{m}}(t) & \text{ if }o=i\text{ and }j<c_{d}\\
0 & \text{ otherwise.}
\end{cases}\\
=\begin{cases}
s_{i}^{j}(t)-\sum_{\mathbf{m}\in\mathbb{M}^{i=j,o=0}\cup\mathbb{M}^{i=j,d=c_{d}}}p_{\mathbf{m}}(t) & \text{ if }o\neq i\text{ and }d\neq i\\
s_{i}^{j-1}(t)-\sum_{\mathbf{m}\in\mathbb{M}^{i=j-1,o=0}}p_{\mathbf{m}}(t) & \text{ if }d=i\text{ and }j>1\\
s_{i}^{j+1}(t)-\sum_{\mathbf{m}\in\mathbb{M}^{i=j+1,d=c_{d}}}p_{\mathbf{m}}(t) & \text{ if }o=i\text{ and }j<c_{d}\\
0 & \text{ otherwise.}
\end{cases}
\end{gather*}

Let $\alpha_{od}^{i=j}(t):=\sum_{\mathbf{m}\in\mathbb{M}^{i=j,o=0}\cup\mathbb{M}^{i=j,d=c_{d}}}p_{\mathbf{m}}(t)$,
$\beta_{o}^{i=j}(t):=\sum_{\mathbf{m}\in\mathbb{M}^{i=j,o=0}}p_{\mathbf{m}}(t)$
and $\gamma_{d}^{i=j}(t):=\sum_{\mathbf{m}\in\mathbb{M}^{i=j,d=c_{d}}}p_{\mathbf{m}}(t)$,
which allows us to express $X_{od}^{i=j}(t)$ more concisely as 
\[
X_{od}^{i=j}(t)=\begin{cases}
s_{i}^{j}(t)-\alpha_{od}^{i=j}(t) & \text{ if }o\neq i\text{ and }d\neq i\\
s_{i}^{j-1}(t)-\beta_{o}^{i=j}(t) & \text{ if }d=i\text{ and }j>0\\
s_{i}^{j+1}(t)-\gamma_{d}^{i=j}(t) & \text{ if }o=i\text{ and }j<c_{d}\\
0 & \text{ otherwise.}
\end{cases}
\]

Function $s_{i}^{j}(t)$ is piecewise differentiable with discontinuous
jumps at time points $t_{1},t_{2},\ldots\in\rho$. For every $o,d\in X$
and every time point $t\in\rho_{od}$, we have
\begin{align*}
s_{i}^{j}(t)= & \sum_{\mathbf{m}\in\mathbb{M}^{i=j}}\lim\limits _{t'\rightarrow t}\,p_{T_{od}^{-1}(\mathbf{m})}(t')\\
= & \lim\limits _{t'\rightarrow t}\,X_{od}^{i=j}(t').
\end{align*}
Using the definition of $s_{i}^{j}$, the derivative of function $s_{i}^{j}$
at every point $t\in\mathbb{R}_{\geq0}\setminus\rho$, $\dot{s}_{i}^{j}(t)=$
$$
\sum_{\mathbf{m}\in\mathbb{M}^{i=j}}\Biggl(\sum\limits _{o,d\in X}\lambda_{od}(t)\cdot p_{T_{od}^{-1}(\mathbf{m})}-p_{\mathbf{m}}\cdot\sum\limits _{o,d\in X}\lambda_{od}\Biggr), $$
which can be rearranged as follows 
\begin{align*}
\dot{s}_{i}^{j}(t)= & \sum\limits _{o,d\in X}\lambda_{od}(t)\sum_{\mathbf{m}\in\mathbb{M}^{i=j}}p_{T_{od}^{-1}(\mathbf{m})}\\
 & -\sum\limits _{o,d\in X}\lambda_{od}(t)\sum_{\mathbf{m}\in\mathbb{M}^{i=j}}p_{\mathbf{m}}\\
= & \sum\limits _{o,d\in X}\lambda_{od}(t)\sum_{\mathbf{m}\in\mathbb{M}^{i=j}}p_{T_{od}^{-1}(\mathbf{m})}\\
 & -s_{i}^{j}(t)\cdot\sum\limits _{o,d\in X}\lambda_{od}(t)\\
= & \sum\limits _{o,d\in X}\lambda_{od}(t)\cdot X_{od}^{i=j}(t)-s_{i}^{j}(t)\cdot\sum\limits _{o,d\in X}\lambda_{od}(t).\\
= & \sum\limits _{o,d\in X,o\neq i,d\neq i}\lambda_{od}(t)\cdot(s_{i}^{j}(t)-\alpha_{od}^{i=j}(t))\\
 & +\sum\limits _{o\in X}\lambda_{oi}(t)\cdot s_{i}^{j-1}(t)-\lambda_{oi}(t)\cdot\beta_{o}^{i=j-1}(t)\\
 & +\sum\limits _{d\in X}\lambda_{id}(t)\cdot s_{i}^{j+1}(t)-\lambda_{id}(t)\cdot\gamma_{d}^{i=j+1}(t)\\
 & -\sum\limits _{o,d\in X}\lambda_{od}(t)\cdot s_{i}^{j}(t)\\
= & \lambda_{i}^{a}(t)\cdot s_{i}^{j-1}(t)\\
 & -(\lambda_{i}^{a}(t)+\lambda_{i}^{d}(t))\cdot s_{i}^{j}(t)\\
 & +\lambda_{i}^{d}(t)\cdot s_{i}^{j+1}(t)\\
 & -\sum\limits _{o,d\in X,o\neq i,d\neq i}\lambda_{od}(t)\cdot\alpha_{od}^{i=j}(t)\\
 & -\sum\limits _{o\in X}\lambda_{oi}(t)\cdot\beta_{o}^{i=j-1}(t)\\
 & -\sum\limits _{d\in X}\lambda_{id}(t)\cdot\gamma_{o}^{i=j+1}(t)
\end{align*}

Recall now the definition of function $q_{i}^{j}$ and let 
\begin{gather*}
Y_{od}^{i=j}(t):=\begin{cases}
q_{i}^{j}(t) & \text{ if }o\neq i\text{ and }d\neq i\\
q_{i}^{j-1}(t) & \text{ if }d=i\text{ and }j>0\\
q_{i}^{j+1}(t) & \text{ if }o=i\text{ and }j<c_{d}\\
0 & \text{otherwise.}
\end{cases}
\end{gather*}
The functions $q_{i}^{j}$ and $s_{i}^{j}$ are equal at $t=0$ and
thus we have $q_{i}^{j}(0)=s_{i}^{j}(0)$. The behavior of $q_{i}^{j}$
at discontinuities is 
\[
q_{i}^{j}(t)=\lim\limits _{t'\rightarrow t^-}q_{i}^{j-1}(t')\quad\forall t\in\rho_{i}^{a}
\]

\[
q_{i}^{j}(t)=\lim\limits _{t'\rightarrow t^-}q_{i}^{j+1}(t')\quad\forall t\in\rho_{i}^{d}
\]
which can be also expressed as: $\forall o,d\in X,\:\forall t\in\rho_{od}$,
\[
\begin{aligned}q_{i}^{j}(t)= & \begin{cases}
\lim\limits _{t'\rightarrow t^-}q_{i}^{j}(t') & \text{ if }o\neq i\text{ and }d\neq i\\
\lim\limits _{t'\rightarrow t^-}q_{i}^{j-1}(t') & \text{ if }d=i\text{ and }j>0\\
\lim\limits _{t'\rightarrow t^-}q_{i}^{j+1}(t') & \text{ if }o=i\text{ and }j<c_{d}\\
0 & \text{otherwise}
\end{cases}\\
= & \lim\limits _{t'\rightarrow t^-}\,Y_{od}^{i=j}(t').
\end{aligned}
\]
The behavior everywhere else is described as: $\forall t\in\mathbb{R}_{\geq0}\setminus\rho,$
\begin{align*}
\dot{q}_{i}^{j}(t)= & \ \lambda_{i}^{a}(t)\cdot q_{i}^{j-1}(t)\\
 & -(\lambda_{i}^{a}(t)+\lambda_{i}^{d}(t))\cdot q_{i}^{j}(t)\\
 & +\lambda_{i}^{d}(t)\cdot q_{i}^{j+1}(t)
\end{align*}
The vector versions of functions $s_{i}^{j}$ and $q_{i}^{j}$ are
denoted as 
\[
\begin{gathered}\mathbf{s}_{i}(t):=[s_{i}^{1}(t),\ldots,s_{i}^{c_{i}}(t)]\\
\mathbf{q}_{i}(t):=[q_{i}^{1}(t),\ldots,q_{i}^{c_{i}}(t)].
\end{gathered}
\]
We will now analyze the behavior of the functions when they are differentiable.
Let $(t_{s},t_{e})$ be an subinterval of positive real line satisfying
$(t_{s},t_{e})\cap\rho=\emptyset$. Then, the evolution of both vector
function $\mathbf{s}_{i}$ and $\mathbf{q}_{i}$ can be described
in a form of matrix differential equation with identical time-dependent
coefficient matrix $Q_{i}$: 
\begin{align}
\mathbf{s}'_{i}(t):= & Q_{i}(t)\cdot s_{i}(t)+\mathbf{b}_{i}(t)\label{eq:s_diff_eq}\\
\mathbf{q}'_{i}(t):= & Q_{i}(t)\cdot q_{i}(t),\label{eq:q_diff_eq}
\end{align}
where $Q_{i}(t):=$ {\footnotesize{}
\begin{gather*}
\left[\begin{array}{ccccc}
-(\lambda_{i}^{d}+\lambda_{i}^{a}) & \lambda_{i}^{d} & 0 & 0 & \cdots\\
\lambda_{i}^{a} & -(\lambda_{i}^{d}+\lambda_{i}^{a}) & \lambda_{i}^{d} & 0 & \cdots\\
0 & \lambda_{i}^{a} & -(\lambda_{i}^{d}+\lambda_{i}^{a}) & \lambda_{i}^{d} & \cdots\\
0 & 0 & \lambda_{i}^{a} & -(\lambda_{i}^{d}+\lambda_{i}^{a}) & \cdots\\
\vdots & \vdots & \vdots & \vdots & \ddots
\end{array}\right].\\
\text{(the argument \ensuremath{t}of functions \ensuremath{\lambda_{i}^{d}} and \ensuremath{\lambda_{i}^{a}} was dropped for brevity)}
\end{gather*}
}and $\mathbf{b}_{i}(t)=[b_{i}^{1}(t),\ldots,b_{i}^{c_{i}}(t)]$ with
$b_{i}^{j}(t):=$
\[
\begin{gathered}-\sum\limits _{o,d\in X,o\neq i,d\neq i}\lambda_{od}(t)\cdot\alpha_{od}^{i=j}(t)\\
-\sum\limits _{o\in X}\lambda_{oi}(t)\cdot\beta_{o}^{i=j-1}(t)\\
-\sum\limits _{d\in X}\lambda_{id}(t)\cdot\gamma_{o}^{i=j+1}(t)
\end{gathered}
\]

Note that we have $\mathbf{b}_{i}^{j}(t)\leq0\ \forall i,j,t.$ The
solution to the homogenous differential equation (\ref{eq:q_diff_eq})
have the form $\mathbf{q}_{i}(t)=\Phi(t,t_s)\mathbf{q}_{i}(t_{s})$, 
$\Phi(t,t_s)=\mathcal{T}e^{\int_{t_{s}}^{t}Q(\tau)d\tau}$, 
where $\mathcal{T}$ is the time ordering operator, 
taking care of the non-commutativity of $Q$ at different times. 
Similarly, the solution to the non-homogenous counterpart (\ref{eq:s_diff_eq}) 
is $\mathbf{s}_{i}(t)=\Phi(t,t_s)\mathbf{s}_{i}(t_{s})+\Phi(t,t_s)\int_{t_s}^{t}\Phi(t',t_s)^{-1}\mathbf{b}_{i}(t')dt'=\Phi(t,t_s)\mathbf{s}_{i}(t_{s})+\int_{t_s}^{t}\Phi(t,t')\mathbf{b}_{i}(t')dt'$. 
The matrix $\Phi(t,t')$ in the integrand is the same that governs the time evolution of the probabilities $\mathbf{q}_i(t)$ in the homogeneus solution, hence it maps non-negative vectors to non-negative vectors. 
The solution $\mathbf{s}_{i}$ is a sum of two terms. Notice that
when $\mathbf{s}_{i}(t_{s})\leq\mathbf{q}_{i}(t_{s})$, then at any
time $t>t_{s}$, the first term has lower or equal value than $\mathbf{q}_{i}$.
Further, since $\mathbf{b}_{i}^{j}(t)\leq0$, the second term is bound
to be zero or negative. 
Consequently, we have 
\begin{gather}
\forall t\in(t_{s},t_{e}),\:t_{s},t_{e}\in\mathbb{R}_{\geq0},\ (t_{s},t_{e})\cap\rho=\emptyset:\nonumber \\
\text{if }\mathbf{s}_{i}(t_{s})\leq\mathbf{q}_{i}(t_{s}),\text{ then }\mathbf{s}_{i}(t)\leq\mathbf{q}_{i}(t).\label{eq:q_upperbounds_s}
\end{gather}

We will now generalize the above result to the entire time domain.
The deterministic rebalancing partitions the time domain into time
points $\tau_{1},\tau_{2},\tau_{3},\ldots\in\rho$ where the functions
$\mathbf{s}_{i}$ and $\mathbf{q}_{i}$ are discontinuous and intervals
$({\tau_{0}=0},\tau_{1}),(\tau_{1},\tau_{2}),(\tau_{2},\tau_{3}),\ldots$
on which $\mathbf{s}_{i}$ and $\mathbf{q}_{i}$ are differentiable.
We will show that $\mathbf{s}_{i}(t')\leq\mathbf{q}_{i}(t')\ \forall t'\in\mathbb{R}_{\geq0}$
by induction over such differentiable intervals.\\
\emph{Base step:} We have $\mathbf{s}_{i}(0)=\mathbf{q}_{i}(0)$ and
thus $\mathbf{s}_{i}(\tau_{0})\leq\mathbf{q}_{i}(\tau_{0})$ holds.\\
\emph{Induction step:} Assume $\mathbf{s}_{i}(\tau_{i-1})\leq\mathbf{q}_{i}(\tau_{i-1})$.
Using the property (\ref{eq:q_upperbounds_s}) we know that $\mathbf{s}_{i}(t)\leq\mathbf{q}_{i}(t)\ \forall t\in[\tau_{i-1},\tau_{i})$.
Observe that if $\mathbf{s}_{i}(t)\leq\mathbf{q}_{i}(t)$, then $X_{od}^{i=j}(t)\leq Y_{od}^{i=j}(t)\ \forall i,j,o,d$
and consequently, we have $\mathbf{s}_{i}(\tau_{i})\leq\mathbf{q}_{i}(\tau_{i})$.

Finally recall that by definition $s_{i}^{j}(t)=P(V{}_{i}(t)=j)$
and $q_{i}^{j}(t)=P(\bar{V}{}_{i}(t)=j)$ and thus we have $P(V_{i}(t)=j)\leq P(\bar{V}{}_{i}(t)=j)$
for all $i\in X$, $j\in\mathbb{M}_{i}$, and $t\geq0$.
\end{proof}

\subsection*{Proof of Corollary~\ref{cor:pf_is_bounded_by_sum_of_qif}:}
\begin{proof}
Recall that 
\[
\dot{p}_{F}(t)=\sum_{o,d\in X}\lambda_{od}(t)\sum_{\mathbf{m}\in\mathbb{M}^{o=0}\cup\mathbb{M}^{d=c_{i}}}\,p_{\mathbf{m}}(t).
\]
 Observe that for particular $o,d\in X$, it holds that 
\[
\sum_{\mathbf{m}\in\mathbb{M}^{o=0}\cup\mathbb{M}^{d=c_{i}}}p_{\mathbf{m}}(t)\leq\sum_{\mathbf{m}\in\mathbb{M}^{o=0}}p_{\mathbf{m}}(t)+\sum_{\mathbf{m}\in\mathbb{M}^{d=c_{d}}}p_{\mathbf{m}}(t).
\]
Consequently, we have 
$$
\dot{p}_{F}(t)\leq \sum_{o,d\in X}\lambda_{od}(t)\left(\sum_{\mathbf{m}\in\mathbb{M}^{o=0}}p_{\mathbf{m}}(t)+\sum_{\mathbf{m}\in\mathbb{M}^{d=c_{d}}}p_{\mathbf{m}}(t)\right),
$$
which implies
\begin{align*}
\dot{p}_{F}(t)\leq & \sum_{o,d\in X}\lambda_{od}(t)\sum_{\mathbf{m}\in\mathbb{M}^{o=0}}p_{\mathbf{m}}(t)\\
 & + \sum_{o,d\in X}\lambda_{od}(t)\sum_{\mathbf{m}\in\mathbb{M}^{d=c_{d}}}p_{\mathbf{m}}(t).
\end{align*}

From Theorem~\ref{thm:decoupled-model-upperbounds-coupled}, we know
$\sum_{\mathbf{m}\in\mathbb{M}^{i=j}}p_{\mathbf{m}}(t)\leq q_{i}^{j}(t)$
and thus 
\begin{align*}
\dot{p}_{F}(t)\leq & \sum_{o,d\in X}\lambda_{od}(t)q_{o}^{0}(t)+\sum_{o,d\in X}\lambda_{od}(t)q_{d}^{c_{d}}(t)\\
\dot{p}_{F}(t)\leq & \sum_{o\in X}q_{o}^{0}(t)\sum_{d\in X}\lambda_{od}(t)+\sum_{d\in X}q_{d}^{c_{d}}(t)\sum_{d\in X}\lambda_{od}(t)\\
\dot{p}_{F}(t)\leq & \sum_{i\in X}\left(\lambda_{i}^{d}(t)q_{i}^{0}(t)+\lambda_{i}^{a}(t)q_{d}^{c_{d}}(t)\right)\\
\dot{p}_{F}(t)\leq & \sum_{i\in X}\dot{q}_{i}^{F}(t).
\end{align*}
From the above upper bound on $\dot{p}_{F}(t)$ and using $p_{F}(0)=0$
and $q_{i}^{F}(0)=0\ \forall i\in X$, we conclude $p_{F}(t)\leq\sum_{i\in X}q_{i}^{F}(t)$. 
\end{proof}

\fi
\end{document}